%% file: sample-sigconf.tex
\documentclass[sigconf]{acmart}

\usepackage{booktabs} 
\usepackage{multirow}
\usepackage{graphicx}
\usepackage{balance}
\usepackage{subfigure}
\usepackage{bm}
\usepackage{amsmath}
\usepackage{amsfonts}
\usepackage{graphicx}

\usepackage[ruled,vlined]{algorithm2e}
\usepackage{algpseudocode}

\newcommand{\squishlist}{
 \begin{list}{$\bullet$}
  { \setlength{\itemsep}{0pt}
     \setlength{\parsep}{3pt}
     \setlength{\topsep}{3pt}
     \setlength{\partopsep}{0pt}
     \setlength{\leftmargin}{1.5em}
     \setlength{\labelwidth}{1em}
     \setlength{\labelsep}{0.5em} } }

\newcommand{\squishlisttwo}{
 \begin{list}{$\bullet$}
  { \setlength{\itemsep}{0pt}
     \setlength{\parsep}{0pt}
    \setlength{\topsep}{0pt}
    \setlength{\partopsep}{0pt}
\setlength{\leftmargin}{2em}
\setlength{\labelwidth}{1.5em}
\setlength{\labelsep}{0.5em} } }

\newcommand{\squishend}{
\end{list}  }

\copyrightyear{2023}
\acmYear{2023}
\setcopyright{rightsretained}
\acmConference[KDD '23]{Proceedings of the 29th ACM SIGKDD Conference on Knowledge Discovery and Data Mining}{August 6--10, 2023}{Long Beach, CA, USA}
\acmBooktitle{Proceedings of the 29th ACM SIGKDD Conference on Knowledge Discovery and Data Mining (KDD '23), August 6--10, 2023, Long Beach, CA, USA}
\acmDOI{10.1145/3580305.3599826}
\acmISBN{979-8-4007-0103-0/23/08}
\begin{document}
\title[Multi-funnel Fresh Content Recommendation]{Fresh Content Needs More Attention:\\ Multi-funnel Fresh Content Recommendation}

\author{Jianling Wang*, Haokai Lu*, Sai Zhang, Bart Locanthi, Haoting Wang, Dylan Greaves, Benjamin Lipshitz, Sriraj Badam, Ed H. Chi, Cristos J. Goodrow, Su-Lin Wu, Lexi Baugher and Minmin Chen}\thanks{* indicates Equal Contribution}
\affiliation{%
  \institution{Google}
}
\email{{jianlingw, haokai, saisaizhang, bnl, haotingwang,lipshitz, srirajdutt, edchi, cristos, sulin, baugher, minminc}@google.com}

\renewcommand{\shortauthors}{Wang, Jianling, et al.}
\begin{abstract}

Recommendation system serves as a conduit connecting users to an incredibly large, diverse and ever growing collection of contents. In practice, missing information on fresh (and tail) contents needs to be filled in order for them to be exposed and discovered by their audience.
We here share our success stories in building a dedicated fresh content recommendation stack on a large commercial platform. To nominate fresh contents, we built a multi-funnel nomination system that combines (i) a two-tower model with strong generalization power for coverage, and (ii) a sequence model with near real-time update on user feedback for relevance. The multi-funnel setup effectively balances between coverage and relevance. An in-depth study uncovers the relationship between user activity level and their proximity toward fresh contents, which further motivates a contextual multi-funnel setup. Nominated fresh candidates are then scored and ranked by systems considering prediction uncertainty to further bootstrap content with less exposure. We evaluate the benefits of the dedicated fresh content recommendation stack, and the multi-funnel nomination system in particular, through user corpus co-diverted live experiments. We conduct multiple rounds of live experiments on a commercial platform serving billion of users demonstrating efficacy of our proposed methods.  

\end{abstract}
%



\maketitle

\input{samplebody-conf}

\balance
\bibliographystyle{ACM-Reference-Format}
\bibliography{sample-bibliography} 


\end{document}

%% file: samplebody-conf.tex
\section{Introduction}
Recommendation systems are heavily relied upon to connect users to the right contents on recommendation platforms. Many of these systems are trained on interaction logs collecting historical user interactions with recommended contents. These interactions however are conditioned on the contents these systems chose to expose to users, which creates a strong feedback loop~\cite{chaney2018algorithmic} resulting in ``rich gets richer'' effect. Fresh uploads, especially those from less popular content providers on the other hand, face a significant barrier to be picked up by these systems and shown to the right users due to \textit{lack of the initial exposure and interaction}. This is commonly known as the item cold-start problem.   

Our goal is to break the feedback loop and create a healthy platform where high-quality fresh contents can surface and go viral as well. To achieve it, we need to \textit{seed the initial exposure of these content to fill in missing knowledge in these systems}, so they can be recommended to the right users. 
Although there is emerging research focusing on facilitating cold-start recommendation~\cite{li2019zero,liu2020heterogeneous,zhu2021learning,zheng2020cold,wang2021sequential} beyond the early work on content-based recommendation \cite{guo2020survey,wang2021survey,zhang2019deep}, how to bootstrap the huge amount of fresh contents (e.g., more than 500 hours of contents are uploaded to YouTube every minute \cite{youtubePress,tubefilter2019}; a new track is uploaded to Spotify every 1.4 seconds \cite{mbw}) on industrial scale recommendation platforms remains under-explored. 

\begin{figure}
\includegraphics[width=0.26\textwidth]{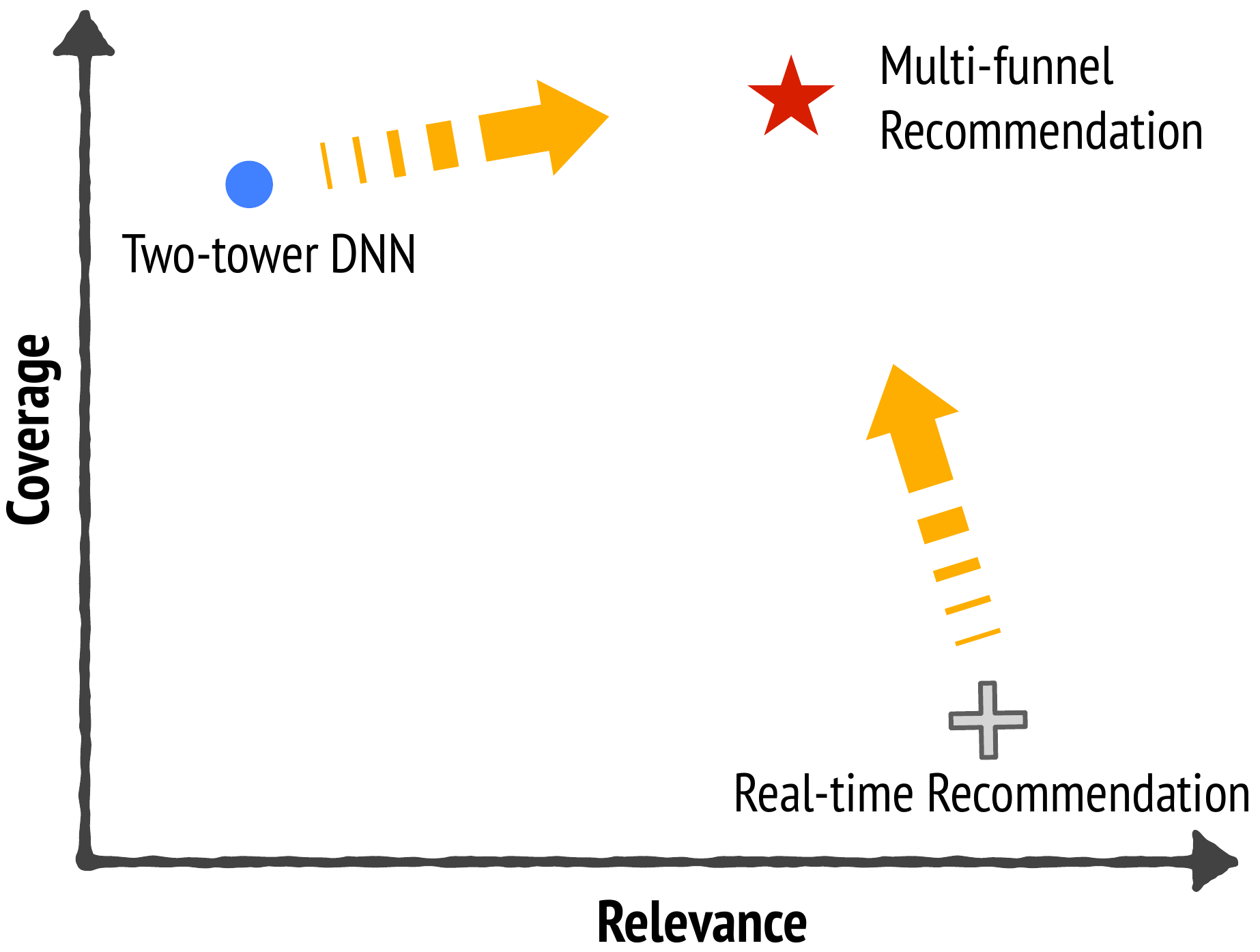}
\caption{Coverage vs. relevance for fresh content recommendation.
}
\label{fig:coverage_vs_relevance}
\end{figure}

More concretely, we aim at building a pipeline to \textit{surface fresh and tail contents} to increase the discoverable corpus\footnote{We refer to the part of corpus that receives more than X (post bootstrapping) positive interactions as the discoverable corpus.} on recommendation platforms. There is a dilemma though: 
a system capable of exploring the full spectrum of the corpus leads to improved long-term user experience~\cite{chen2021exploration,jadidinejad2020using}; however as less certain contents are recommended, fresh content recommendation often comes at a cost to short-term user experience, which we would like to mitigate.  
To balance the short-term and long-term user experience in the fresh content recommendation pipeline, we measure its effectiveness on two dimensions: (i) \textbf{coverage} to examine if the pipeline can get more unique fresh contents exposed to the users, and (ii) \textbf{relevance} to examine whether the system is connecting users with fresh contents that are interesting to the users.


Designing the fresh content recommendation stacks still faces many unknowns: (i) \textit{how to position the fresh content recommendation stack w.r.t. existing recommendation stacks?} We choose to build a dedicated fresh content recommendation stack that is relatively separated from the rest of the recommendation stacks which focus on relevance, and are more susceptible to strong feedback loop; (ii) \textit{what components are needed in the stack?} We design the stack to include a nomination system, graduation filter and a ranking system to score and rank the candidates; (iii) \textit{how to balance between coverage and relevance?} A system randomly recommending fresh contents will reach maximum coverage at the expense of low relevance and impacted user experience. To achieve a balance, we built a multi-funnel nomination system that diverts user requests between a model with high coverage and one with high relevance (see Figure \ref{fig:coverage_vs_relevance}); (iv) \textit{how to model contents with little to no prior engagement data?}  We rely on a two-tower DNN model that leverages content features for generalization to bootstrap the initial distribution, and a sequence model that updates on near real-time user feedback to quickly find audience for the high-quality content; and (v) \textit{how to measure the benefits of fresh content recommendation?} We adopt a user-corpus co-diverted experiment framework to measure the effect of the proposed system in increasing the discoverable corpus.
We organize our paper as follows and highlight our key contributions:
\squishlist
\item In Section \ref{sec:overview}, we design a dedicated multi-stage fresh content recommendation stack with nomination, filtering, scoring and ranking components to effectively bootstrap cold-start items; We showcase its values in improving corpus coverage, discoverable corpus and content creation on a commercial recommendation platform serving billions of users.
\item In section \ref{sec:multi-funnel-section}, we design a multi-funnel nomination system within the dedicated fresh content recommendation stack that combines a model with high generalization capability and one with near real-time update to effectively balance between coverage and relevance in fresh recommendation. We conduct a series of live experiments on the same platform to demonstrate the efficacy of the multi-funnel nomination system and summarize the results in Section \ref{sec:exp}.
\item In Section \ref{sec:contextual}, we extend the system to divert user requests to these two models based on request context to further improve fresh recommendation efficiency. 
\squishend


\section{Fresh Content Recommendation}
\label{sec:overview}
\smallskip
\noindent\textbf{Pipeline Setup.} 
We first introduce a dedicated fresh recommendation stack to surface relevant fresh and tail contents for users in one of the largest commercial recommendation platform serving billions of users.
The production recommender system has multiple stages \cite{covington2016deep,ma2020off}, in which the first stage includes multiple retrieval models to nominate candidates from the overall corpus, the second stage scores and ranks the candidates, while the last stage packs the selected candidates for diversity and different business goals. Fresh and tail items however are hard to be discovered by the system due to the closed feedback loop~\cite{jiang2019degenerate,chaney2018algorithmic}. To surface such contents, 
we dedicate one (floating) slot to fresh ($\leqslant X$  days old) and tail (with less than $Y$ positive user interactions) contents. The rest of the slots are still filled with the production recommendation system. 
The pipeline for this dedicated fresh  recommendation system is illustrated in Figure \ref{fig:fixed_slot}, with each component described as follows:

\begin{figure}
\includegraphics[width=0.49\textwidth]{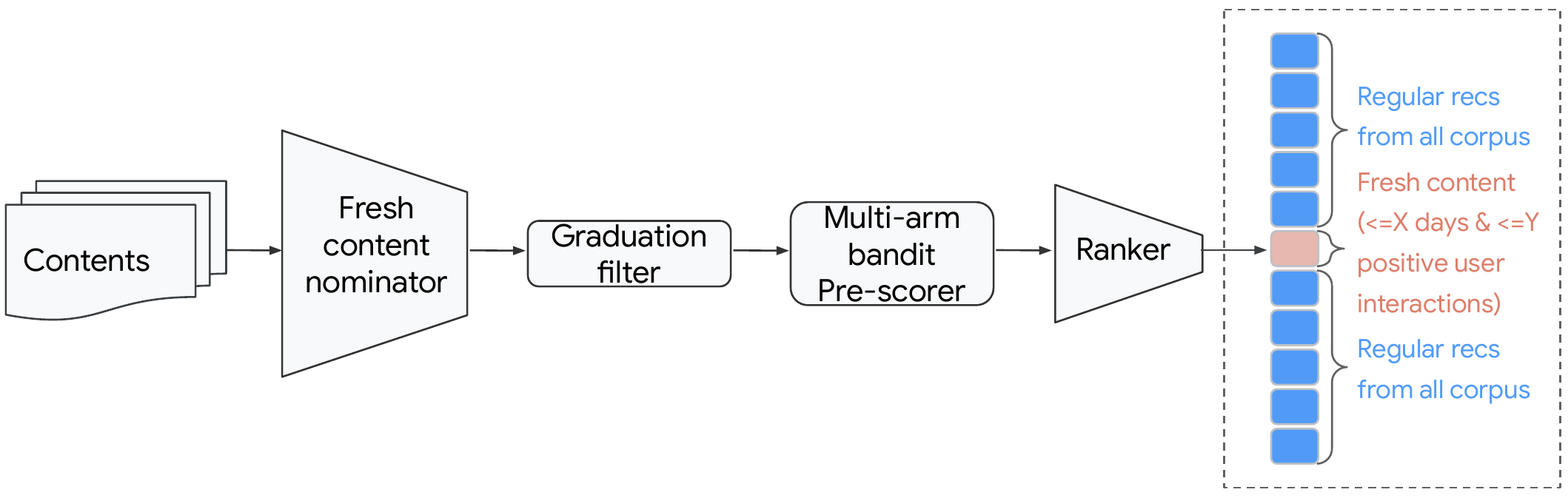}
\caption{A dedicated fresh content recommendation stack.
}

\label{fig:fixed_slot}
\end{figure}

\squishlist
\item {\textbf{Fresh content nominator}}. A key challenge in nominating relevant fresh and tail contents lies in the lack of user interaction data on them. To overcome the cold-start item recommendation problem, we adopt a two-tower content-based recommendation model \cite{yi2019sampling,huang2013learning,rybakov2018effectiveness}: a query tower is used to learn user representation based on the consumption histories; and a candidate tower is used to learn item representation based on item features. The model is trained to predict positive user interactions from historical data, and fortunately through item features, it is able to generalize to items with zero or very little user interactions. At serving time, we rely on a multiscale quantization approach for fast approximate Maximum Inner Product Search \cite{guo2016quantization, wu2017multiscale} and retrieve the top-50 fresh candidates efficiently.  

\item {\textbf{Graduation filter}}. Once fresh contents accumulate initial interactions, they can be readily picked up by the main recommendation system and shown to the right users in other slots. There is diminishing return to continue exploring these items in the dedicated slot, and the impression resource can be  
 saved and re-allocated to other potential fresh and tail contents. We adopt a graduation filter that removes contents which have been consumed by users for at least $n$ times at real-time\footnote{We chose the threshold based on the number of positive interactions needed to be distributed by the main recommender system.}. 

\item {\textbf{Ranking}}. 
Once nominated candidates passed through the filter, we rank them through two components:
a real-time pre-scorer and the ranker shared with the main system. The pre-scorer is lightweight and is able to react to users' early feedback in real-time; The top-10 candidates chosen by the pre-scorer will then pass through the ranker, which is a high-capacity deep neural network~\cite{covington2016deep} with much better accuracy but longer latency, to assess the relevance of the selected candidates and return the top-1. Specifically, the lightweight pre-scorer implements a real-time Bernoulli multi-arm bandit~\cite{slivkins2019introduction} where each arm corresponds to one content. The reward of each arm estimates the good CTR of a fresh content, i.e., click-through rate conditioning on users spending at least 10 seconds after the click, and follows a Beta posterior distribution:
\begin{equation}
  r_i \sim Beta(\alpha_0 + x_i, \beta_0 + n_i - x_i),
  \label{eqn:beta_bandit}
\end{equation}
where $r_i$ is the reward for arm $i$ with a prior $Beta(\alpha_0, \beta_0)$ for some parameters $\alpha_0, \beta_0 \in \mathbb R_{> 0}$, $x_i$ and $n_i$ represent the total number of interactions and impressions for arm $i$ at real-time, respectively. In practice, we estimate the global prior $\alpha_0$ and $\beta_0$ through maximum likelihood estimate on fresh items with at least 100 impressions. At serving time, we adopt Thompson Sampling to generate a sampled reward from Equation~\ref{eqn:beta_bandit} for each candidate, and return the top 10 with the highest rewards, with the final candidate determined by the ranker.
\squishend


This ``dedicated slot'' enables us to measure various fresh content recommendation treatments with more control and flexibility. Compared to ``full-system'' fresh content recommendation (i.e., enable more exploration within the current recommendation stacks and potentially affect every single slot), setting up the ``dedicated slot'' has several advantages: 
(i) \textbf{Deliverable}.
By setting aside a dedicated slot for fresh content only, they can now more easily reach their target audience once uploaded.  
These contents would otherwise face severe competition from the head and more popular contents due to popularity bias in the main recommendation system. 
(ii) \textbf{Measurable}. 
With a dedicated slot and a much simpler pipeline, one can more easily set up different treatments and measure the corpus effect accurately through the following proposed user-corpus diverted experiment. 
(iii) \textbf{Extensible}. The fresh recommendation stack can be easily expanded by extending the single slot treatment to multiple ones, allowing sustainable discoverable corpus growth.


\begin{figure}

\includegraphics[width=0.48\textwidth]{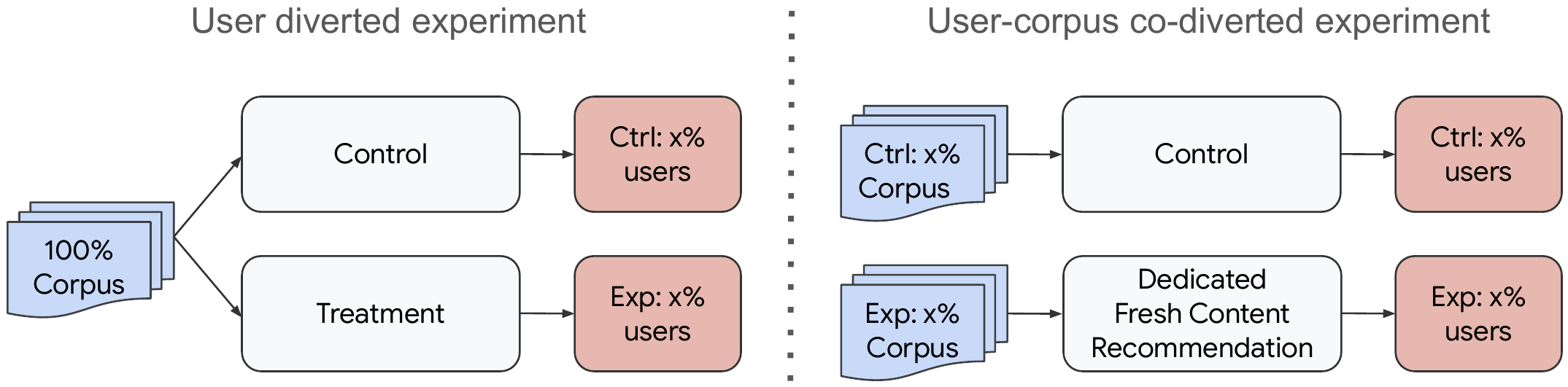}
\caption{User diverted vs. User corpus co-diverted experiment diagrams.}

\label{fig:user-corpus_exp}
\end{figure}

\begin{figure*}[!t]
    \graphicspath{{figures/}}
    \scalebox{0.99}{
    \subfigure[]
    {
    \includegraphics[height=0.21\textwidth]{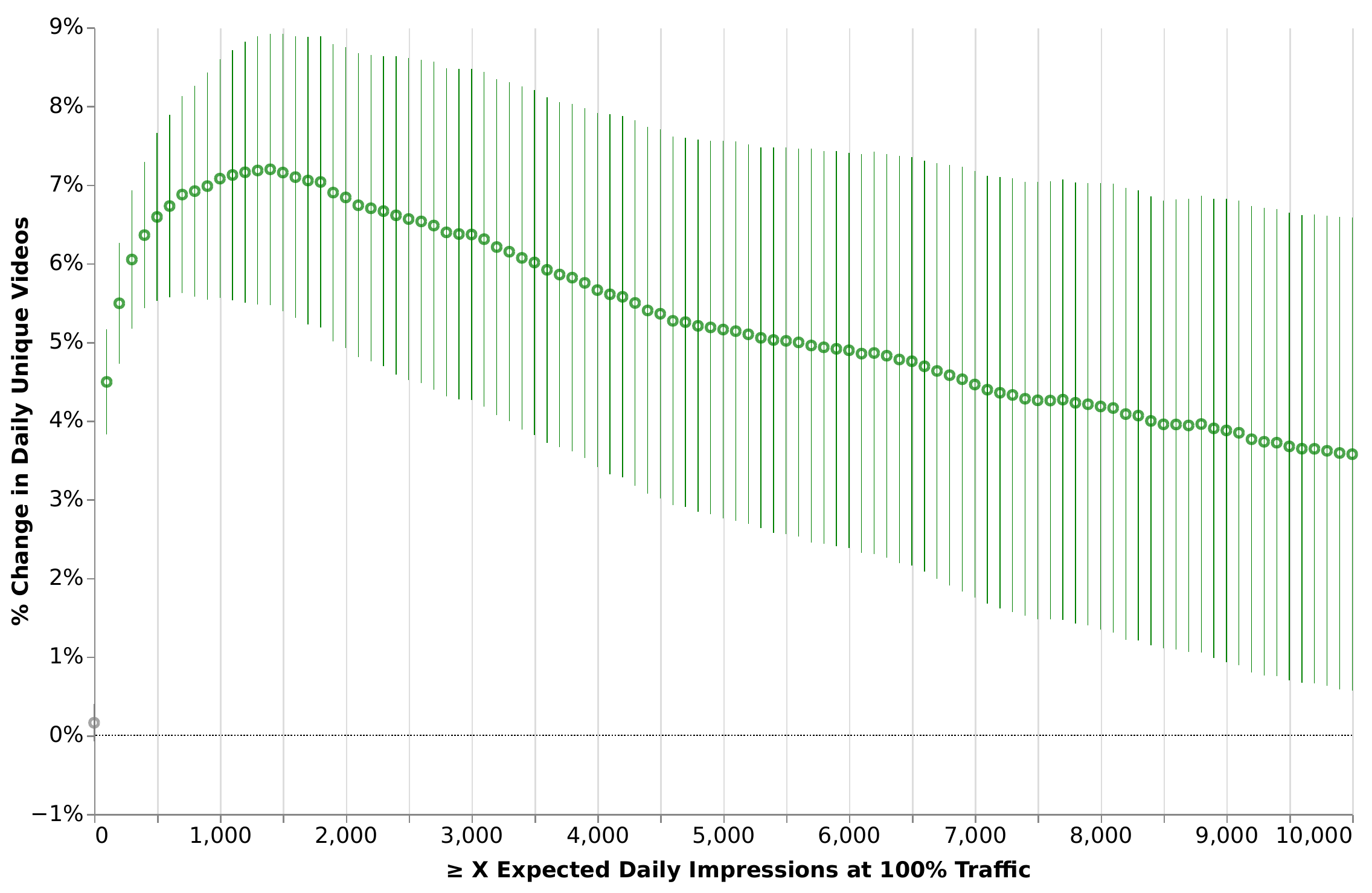}
    }
    \hspace{0.0cm}
    \subfigure[]
    {
    \includegraphics[height=0.21\textwidth]{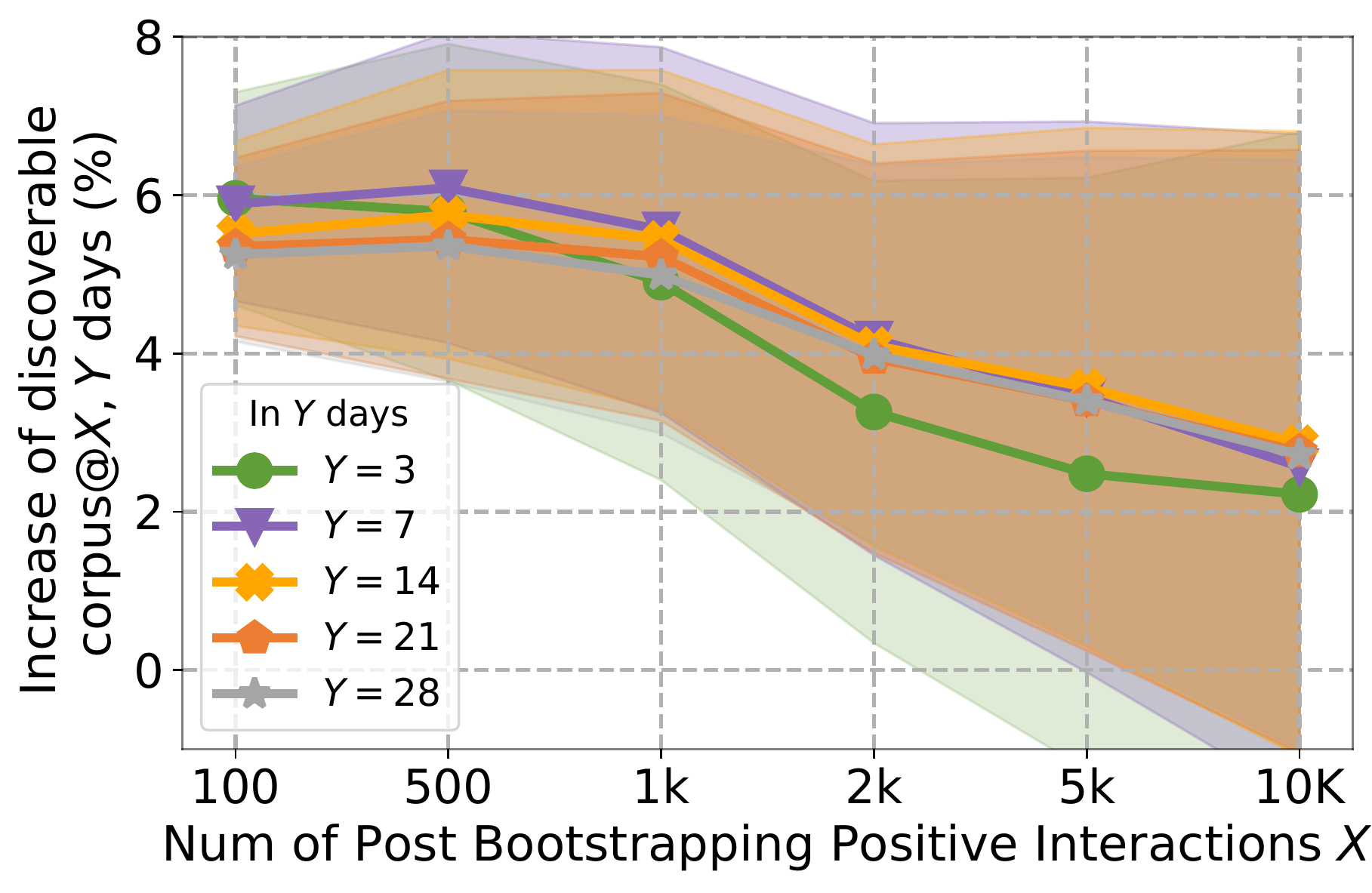}
    }}
    \subfigure[]
    {
    \includegraphics[height=0.21\textwidth]{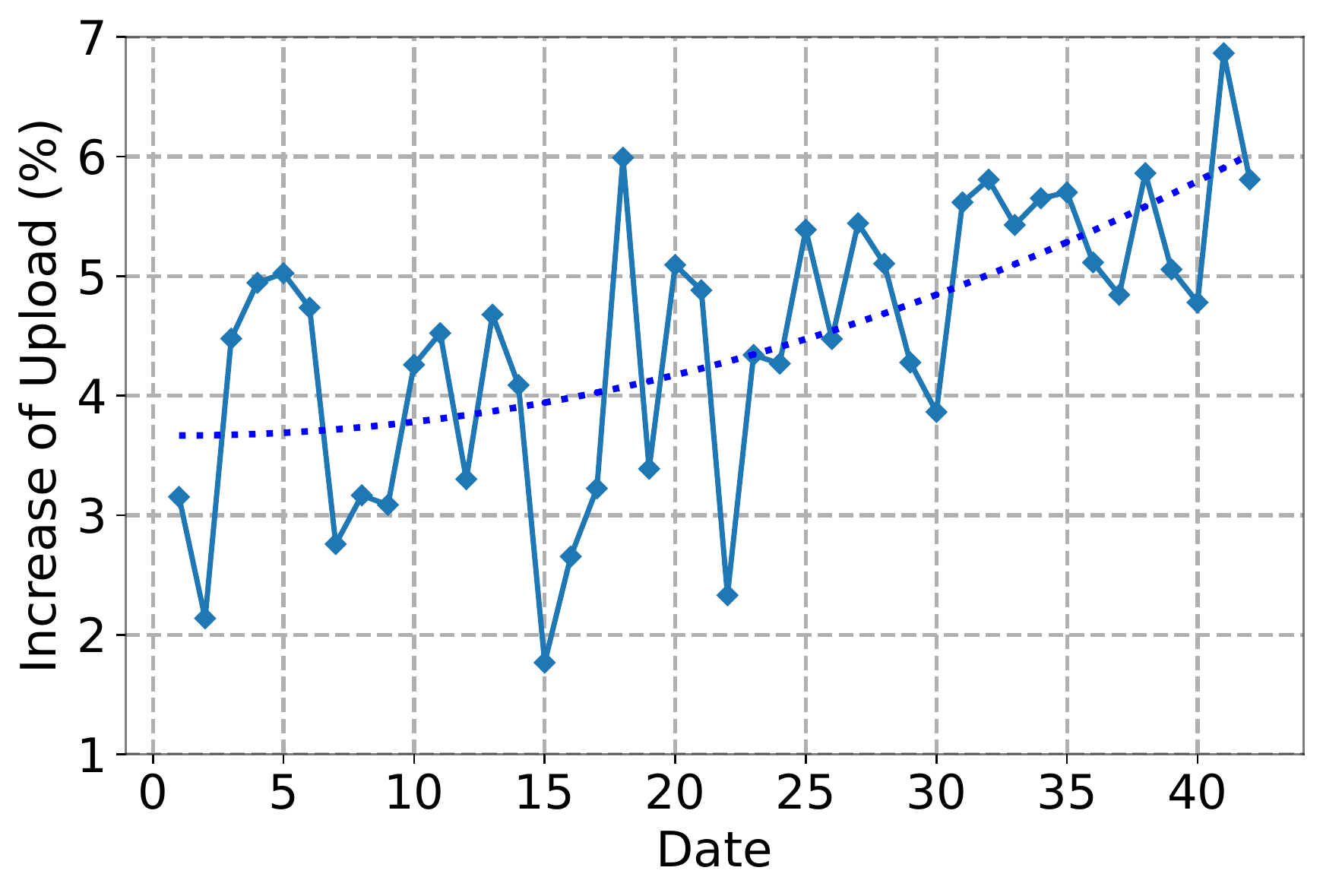}
    }
    \caption{With the dedicated slot for fresh content recommendation, we show (a)  improvement of daily unique impressed contents (with 95\% confidence level) across different impression thresholds;  (b) improvement of number of contents receiving X post graduation good clicks in Y days post graduation; and (c) the dedicated fresh content recommendation system encourages more uploads from content providers in the platform and there is a rising trend as the experiment continues.}%
    \vspace{-0.05in}
    \label{fig:preliminary1}
    \label{fig:explain1}
    \label{fig:viz1}
    \label{fig:explain2}
\end{figure*} 



\smallskip
\noindent\textbf{User corpus co-diverted experiment.} In traditional A/B testing, we usually adopt a user-diverted setup as shown in Figure \ref{fig:user-corpus_exp} (left), in which users are randomly assigned to the control and treatment groups, and receive the corresponding recommendations from the entire corpus. One can compare user metrics such as CTR and dwell time between the two arms, and measure the effect of the treatment from the user perspective. However, since the two arms share the same corpus, the user-diverted setup is unable to measure any treatment effect on the corpus due to treatment leakage. For example, a fresh item receiving exposure through the fresh  recommendation  stack in the experiment can also appear in the control arm. To overcome such leakage,
we adopt a user-corpus-diverted A/B testing setup (in Figure \ref{fig:user-corpus_exp} (right)), in which we  first set aside x\% corpus for the control arm and another non-overlapping x\% corpus for the treatment arm. Then users are assigned proportionally to different arms, meaning that users in the control arm can only receive contents from the control arm corpus, and users in the treatment arm can only receive contents from the treatment arm\footnote{Note that we often further restrict the diversion to be by content providers. In other words,  items belonging to the same provider are assigned to the same arm, to avoid treatment leakage between them.}. Since the size of user and corpus is in proportion, i.e, x\% users are exposed to x\% corpus during experiments, the effect of the treatment measured in the experiment phase is consistent with full deployment when 100\% of the users are exposed to 100\% corpus.

\smallskip
\noindent\textbf{Performance Evaluation Metrics.} We use the following \emph{corpus metrics} to measure the performance of the system in recommending personalized fresh contents, from both the coverage and relevance dimensions: 
\squishlist
\item \textbf{Daily Unique Impressed Contents at $K$} (DUIC@K) is a corpus metric that counts the number of unique contents receiving $K$ impressions daily. We focus on the low end, i.e., relatively small $K$s, to quantify the \emph{coverage} change.
\item \textbf{Fresh Content Dwell Time} (DwellTime) measures time users spent on the impressed fresh content. Longer dwell time indicates the system can infer users' preference on fresh contents more accurately, thus achieving higher \emph{relevance}.
\item \textbf{Number of content receiving X (post bootstrapping) positive interactions in Y days} (Discoverable Corpus@X,Ydays) measures the long-term benefits of fresh content recommendation in incubating quality fresh contents to success. By post bootstrapping, we do not count interactions the item received from the dedicated fresh recommendation stack. A larger discoverable corpus indicates that the system can uncover and seed the success of more worthy content i.e., the ones that can attract positive interactions and reach viral on their own after exiting the dedicated slot.
\squishend
Meanwhile, to ensure that the newly introduced fresh content recommendation does not sacrifice too much short-term user engagement, we also look at a \emph{user metric} measuring the \textit{overall user dwell time on the platform}.

\subsection{Values of Fresh Recommendation}
\label{sec:motivation}
We first conduct user corpus co-diverted live experiments on a commercial recommendation platform serving billions of users to measure the benefits of setting up the proposed fresh content recommendation stack over a one-month period. In these experiments, the users in the control arm are shown only recommendations generated by the main recommendation system. In the treatment arm, a dedicated slot is reserved to show recommendations from the fresh recommendation stack while the other slots are filled by the same recommendation system as the control arm. 
We make the following observations:
\squishlist
\item\textit{Corpus coverage is improved.} Figure \ref{fig:explain1}(a) plots the corpus coverage metric -- Daily Unique Impressed Contents at K (DUIC@K). One can observe a consistent increase of corpus coverage in $4\%\sim7.2\%$ across different K values. As an example, there are $7.2\%$ more unique contents receiving more than 1,000 impressions daily in the treatment arm due to the dedicated fresh  recommendation stack compared with the control arm. As expected, the increase is more pronounced at lower $K$ values. 
\item \textit{A larger corpus is discovered by the users.} 
In Figure \ref{fig:explain1}(b), we plot the discoverable corpus metric, which measures the improvement in the number of contents receiving $X$ post bootstrapping positive interactions in $Y$ days.
Again one can observe a consistent improvement for a range of values for $X$ (from 100 to 10K), and $Y$ (from 3 days to 28 days). In other words, with the initial exposure and interactions seeded in the dedicated fresh recommendation stacks, more unique number of contents are now being recommended by the main recommendation system and being discovered by the users as a result. The improvement also re-assures that the fresh recommendation stack not only increases the corpus coverage, but also bootstraps worthy contents that can survive on their own quality and relevance after the treatment.
With a larger discoverable corpus, more users will now be able to find contents centering around their specific preferences in topic, style, content provider etc., leading to better user experience and a healthier platform~\cite{chen2021values}. Although it takes time for a fresh upload to be picked up by the main recommendation system and get discovered, we  do find the numbers tend to converge after 7 days, even for those high end $X$ values. Thus, we use discoverable corpus@X,7days as the main discoverable corpus metric in our live experiment to reach conclusion sooner. 


\item\textit{Content providers are encouraged to upload more contents.} Figure \ref{fig:explain2}(c) plots the increase in content uploads in the treatment arm with the dedicated fresh content recommendation stack, compared to the control arm. Throughout the one-month experiment period, a consistent improvement can be observed. Additionally, we note a rising trend as the experiment continues. 
By setting aside the dedicated slot focusing on fresh content recommendation, content providers are encouraged to upload more contents as their newly uploaded items receive more exposures and gain traction from users.


\begin{figure*}
\vspace{-0.1in}
    \graphicspath{{figures/}}
    \centering
    \scalebox{0.99}{
    \subfigure[]
    {
    \hspace{-0.1cm}
    \includegraphics[width=0.32\textwidth]{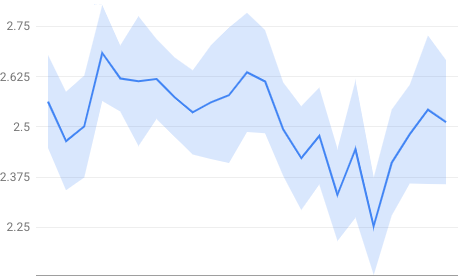}
    }
    \hspace{-0.1cm}
    \subfigure[]
    {
    \includegraphics[width=0.32\textwidth]{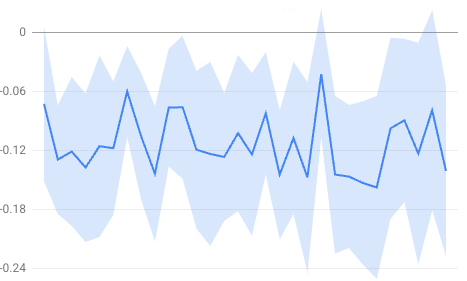}
    }
    \hspace{-0.1cm}
    \subfigure[]
    {
    \includegraphics[width=0.32\textwidth]{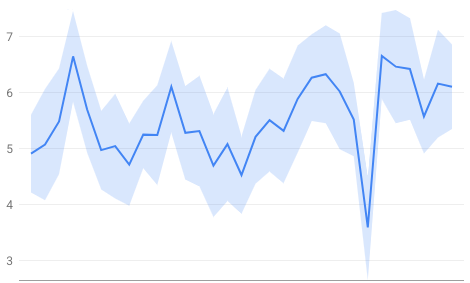}
    }
    }
    \vspace{-0.05in}
    \caption{(a) Fresh 7-day positive interactions (\% change). (b) Overall user dwell time on the platform (\% change). (c) User dwell time for small content providers (\% change).}%
    \vspace{-0.05in}
    \label{fig:explain4}
\end{figure*} 

\item\textit{More fresh content is consumed by the users with a minor impact on short-term user engagement.} Figure \ref{fig:explain4}(a) demonstrates a significant increase in the number of fresh 7-day positive interactions by $+2.52\%$ on average. In Figure \ref{fig:explain4}(b), we find that the overall user dwell time spent on the platform drops slightly by $-0.12\%$. However, as shown in Figure \ref{fig:explain4}(c), the user dwell time for small content providers, those with less than a certain number of subscribers, is increased significantly by $+5.5\%$. 
The trade-off is desirable considering the long-term benefits of a larger discoverable corpus and more active content providers, as explained above.

\squishend



With the established dedicated slot setup for treatment and user corpus co-diverted experiment framework for measurement, we now look into solutions to further increase efficiency of the fresh  recommendation stack.

\section{Multi-funnel Fresh Content Recommendation}
\label{sec:multi-funnel-section}
Different from the main recommendation system, the dedicated fresh recommendation stack focuses on fresh items which have not accumulated enough interaction signals for learning. In this section, we describe how we extend the fresh content nominator in the stack to achieve both high coverage and high relevance when nominating fresh contents. The technique can easily generalize to other components in the pipeline. We center the discussion around the following questions: 
\textbf{RQ1} For fresh contents with no or extremely limited interactions, how to effectively infer their relevance to users and bootstrap these contents? 
\textbf{RQ2} After accumulating some initial interaction feedback, how to quickly take advantage of the limited real-time user feedback to amplify worthy contents? 
\textbf{RQ3} How to balance between content generalization and real-time learning, and reduce the user cost for fresh recommendation, so that we can achieve both high coverage and high relevance when recommending fresh contents?


\subsection{Content Generalization}
\label{sec:content_generalization}
Most of the collaborative filtering-based recommendation models \cite{he2017neural,wang2019neural,yang2020mixed} rely on a factorization backbone to obtain a set of user and content/item ID embeddings based on historical user interactions or ratings. The learned embeddings are later used to infer users' preference on any contents. Due to the long-tailed distribution of items, there are not enough engagement labels for the newly uploaded contents to learn an informative ID embedding \cite{lin2022quantifying,beutel2017beyond}. In practice, without appropriate treatment, the small number of labels from fresh and tail contents are usually ignored by the model and become training noise. These contents as a result are rarely surfaced to the right users. 

The main challenge lies in the lack of existing interaction labels between users and these fresh uploads. 
A solution is to use a content provider-aware recommender which can bootstrap new uploads produced by a provider the user is familiar with or has already subscribed to, but not others. To overcome the scarcity of interaction labels, we rely on content-based features to characterize the new uploads as in \cite{wang2021exploring,lee2017large,van2013deep}. These content features allow the models to generalize from the abundant engagement labels from popular and established contents to the fresh uploads of similar features.

The  model structure follows the two-tower architecture as in \cite{yi2019sampling}, which utilizes a user and an item/content tower, to encode the information from users and items respectively. The dot product between these two towers are learned to predict the historical preference between an user and an item. The model by itself however is still subject to popularity bias. To adapt the model for fresh content recommendation, we made the following changes: 1) we drop item ID completely in the item tower to prevent the model from memorizing historical preferences on individual items; 2) we also exclude any features indicative of historical popularity of an item, e.g., number of impressions, positive engagements, to reduce popularity bias in the item tower. In other words, only meta features that can generalize between popular and newly uploaded contents are included in the item tower for learning. 

\begin{figure}[b]
\includegraphics[width=0.35\textwidth]{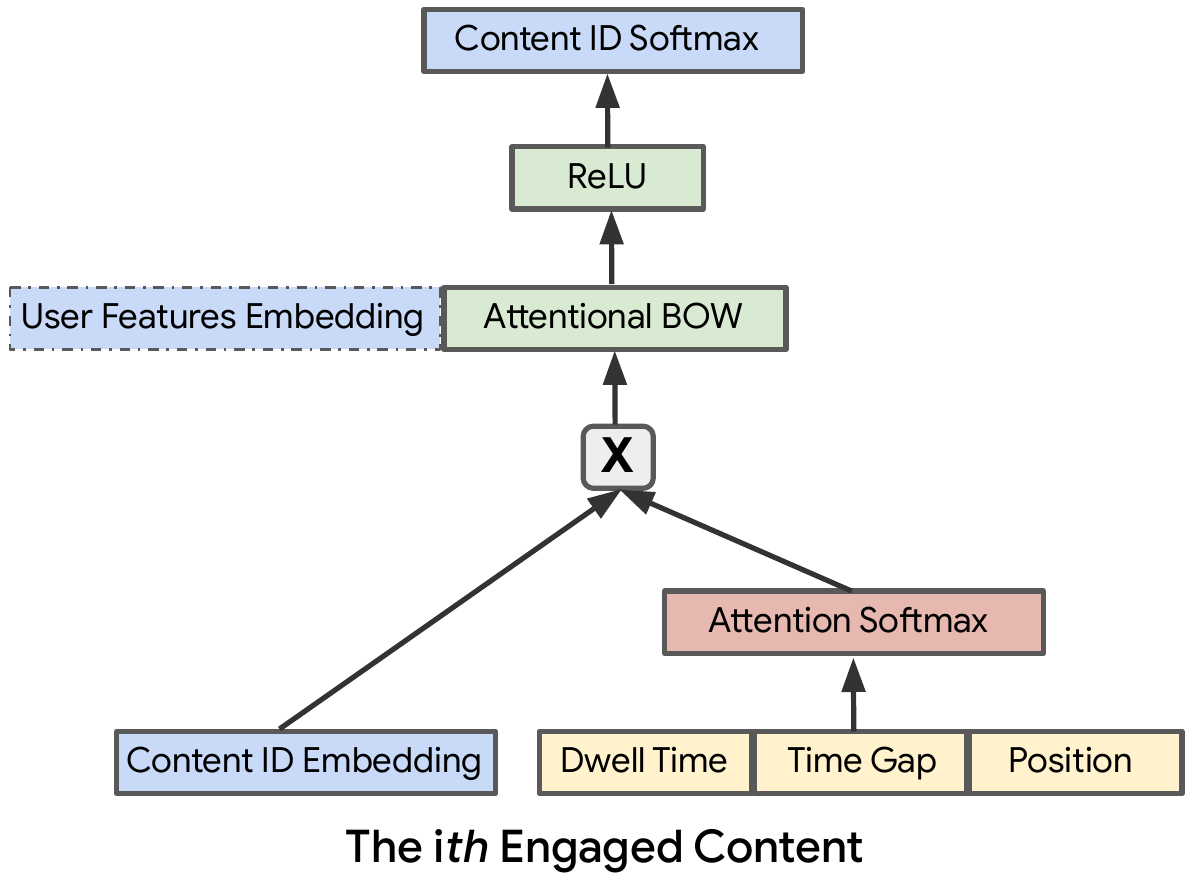}
\caption{Real-Time Sequence Model for Recommendation. }
\label{fig:online_rec}
\end{figure}

We conduct an online A/B testing to measure the impact of the aforementioned changes on improving coverage and relevance of fresh recommendation. The control arm runs a nominator model with the two tower structure as explained in section~\ref{sec:overview}, including all the meta features associated with an item in the item tower. The treatment arm runs the exact same model, but excluding item ID and popularity features from the item tower. 
We observe that, by removing item ID embeddings and item popularity features, the corpus coverage metric, i.e., Daily Unique Impressed Content at 1,000 is increased by 3.3\% with a 95\% confidence interval of be $[3.0\%, 3.7\%]$.  The fresh content dwell time also increases by 2.6\%. The control model can rely on item/content ID embeddings to memorize interaction labels from popular and established contents, but perform poorly on fresh uploads.  The treatment model, however, relies on content features to characterize user preference on these popular and established contents, and applies that learning to fresh uploads with similar features, resulting in improved relevance for fresh recommendation.



\smallskip
\noindent\textbf{Content Features in Used.} The content features used in both the control and treatment arm include multiple categorical features of different granularity derived from the content itself and describe the semantic topic, taxonomic topic and language of the content. We also include average rating to filter out the low-quality contents.

\subsection{Real-Time Learning}
While a nominator heavily relying on content features for generalization is effective in bootstrapping fresh contents with little to no user interaction data, it lacks the memorization capability to \textit{react to users' initial feedback quickly}. Such rapid responsiveness is indeed essential as (i) we often do not have all the features needed to fully characterize the content and reflect the quality of a newly uploaded content; (ii) prompt reaction to initial user feedback can help correct mistakes in early distribution of low-quality or less relevant fresh contents to reduce cost, as well as quickly redistribute and further amplify high-quality and relevant fresh contents to other audiences with similar interests, to further enhance discoverable corpus growth and content provider incentives to upload more. 
This calls for a \textit{near real-time} nominator that can ingest data as new interaction data comes along in a streaming fashion. 

To build such a near real-time nominator, we propose to (i) utilize the near real-time user interaction data for training and (ii) bring up a low-latency personalized retrieval model. We start with minimizing runtime of different components in building this nominator, i.e., data generation, model training and model pushing, so that the end-to-end latency between a user interaction happens and a model (updated with that interaction) is used for serving is a couple of hours. Note this is a \textit{significant latency improvement} over any of the existing recommendation models in the main recommendation stack, which has an end-to-end latency span 18 to 24 hours or even days. 
The data generation job collects the most recent 15 minutes user interactions on fresh and tail contents, which is used as the labels to train the retrieval model.

The retrieval model is trained on the task of predicting the next item a user will interact with given the historic interactions on the platform. The architecture again uses a two-tower structure, where the user/query tower encodes user interaction sequences, and the item tower uses simple ID embeddings as well as categorical features, as shown in Figure \ref{fig:online_rec}. To reduce the training time, we design the model with a simple architecture. The user state is represented as a weighted sum of embeddings of content IDs he or she interacted with recently, concatenated with user query features. Specifically, we apply attention \cite{vaswani2017attention} to improve the user state representation. For a given user with the most recent $n$ (i.e., 500) engaged contents $[V_1, V_2, V_i,...,V_n]$, instead of a naive average, we adopt a weighted sum of the last $n$ interactions of the following form to obtain the user representation $U$:
\begin{displaymath}
  U = \sum_{i=1}^n w_i*Embedding(V_i),
  \label{eqn:user_representation}
\end{displaymath}
where the weight $w_i$ for each content $V_i$ is the normalized softmax weight in $[0,1]$, derived from item features with  
\begin{displaymath}
 w_i = \mbox{softmax}\left(f(dwelltime(V_i), timegap(V_i), position(V_i)\right),
  \label{eqn:w_i}
\end{displaymath}
in which \textit{dwelltime} captures the time the user spent on item $V_i$, and \textit{timegap} captures the time gap between the interaction happens and the current request time. These features are quantized and encoded as discrete embeddings. $f$ then learns a mapping from these embeddings to a scalar to come up with the final weight $w_i$ for the historical interaction on item $V_i$. 
Such a weighted score $w_i$ will be able to highlight the content closer to the current request with longer dwell time when aggregating the historical interactions. 
In our experiment, we find that changing the aforementioned simple attention design with more complex sequence models such as RNNs did not improve the performance, we thus kept the simple weighted embedding for minimal model complexity. The model warm-starts from its previous checkpoint, and trains on the latest 15-minute logs for around an hour in every training round. After that it is published to servers in different data centers to serve the live traffic. At serving time, we again rely on a multiscale quantization approach for fast approximate Maximum Inner Product Search \cite{guo2016quantization, wu2017multiscale} and retrieve the top-50 fresh candidates efficiently.   

\smallskip
\noindent\textbf{Category-centric Reweighting.} To enable fast memorization on the early user feedback on these fresh contents, we include item ID embeddings in the item tower for the real-time nominator. Even among the fresh uploads, the interaction data can vary greatly in pattern: some can accumulate thousands of interaction data in minutes, while others might only see a handful of interactions in that 15-minute log. A model only relying on ID embeddings might run the risk of over-indexing on the "head" contents among the fresh uploads due to the imbalanced interaction data. 
To overcome this issue, we also include some content features mentioned in Section \ref{sec:content_generalization} to characterize the items. However, many categorical features also fall into long-tail distribution. Some categorical features are broad and apply to a large number of items such as ``music'' while others are more specific and informative such as ``Lady Gaga Enigma + Jazz $\&$ Piano''. IDF-weighting where we weigh a feature by inverse of its popularity in the overall item corpus, is introduced so the model can focus on learning these more specific content features for generalization while ignoring the broad ones. 



\subsection{Low-funnel VS Middle-funnel Contents}
\label{sec:multi-funnel}
\begin{figure}[h]
\includegraphics[width=0.4\textwidth]{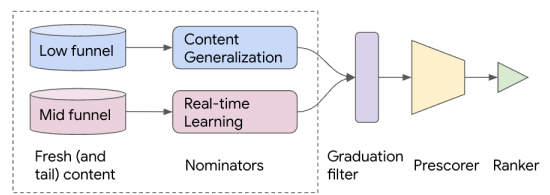}
\vspace{-0.05in}
\caption{A multi-funnel nomination system.}
\label{fig:multi_funnel_system}
\end{figure}

As discussed, trade-off exists in our goals of achieving high coverage and relevance in recommending fresh content. The corpus of the dedicated fresh  recommendation stack, indeed can be further categorized into two parts: (i) \textit{low-funnel} content with very limited or even zero interactions, and (ii) \textit{middle-funnel} content which have collected a few initial interaction feedback through content generalization. For low-funnel content, real-time learning framework loses its predictive power, and generalization on these contents is in great need. On the other hand, for middle-funnel content, the early feedback can power the training of the real-time nomination system, allowing better personalization and relevance.  
Instead of trying to achieve both good generalization and real-time learning with a single nominator\footnote{It is in general hard for a single nominator to most efficiently achieve both. For example, good generalization might require ablating ID embeddings, while real-time learning might need ID embeddings for fast memorization to react to real-time feedback.}, we  decompose the task by deploying different nominators for different funnels (as shown in Figure \ref{fig:multi_funnel_system}): one nominator with good generalization performance targeting low-mid funnel; and another one focusing on adapting to user feedback quickly, targeting mid-funnel (with reasonable amount of user feedback to start with). We adopt the idea of serving two or more recommenders simultaneously to gain better performance with fewer of the drawbacks of any individual one~\cite{burke2002hybrid}. And we will discuss how we determine when to transition a low-funnel content to middle-funnel in such a hybrid system in Section \ref{sec:different_p}. 



\begin{figure*}
\vspace{-0.1in}
    \graphicspath{{figures/}}
    \centering
    \scalebox{0.99}{
    \subfigure[]
    {
    \hspace{-0.1cm}
    \includegraphics[width=0.36\textwidth]{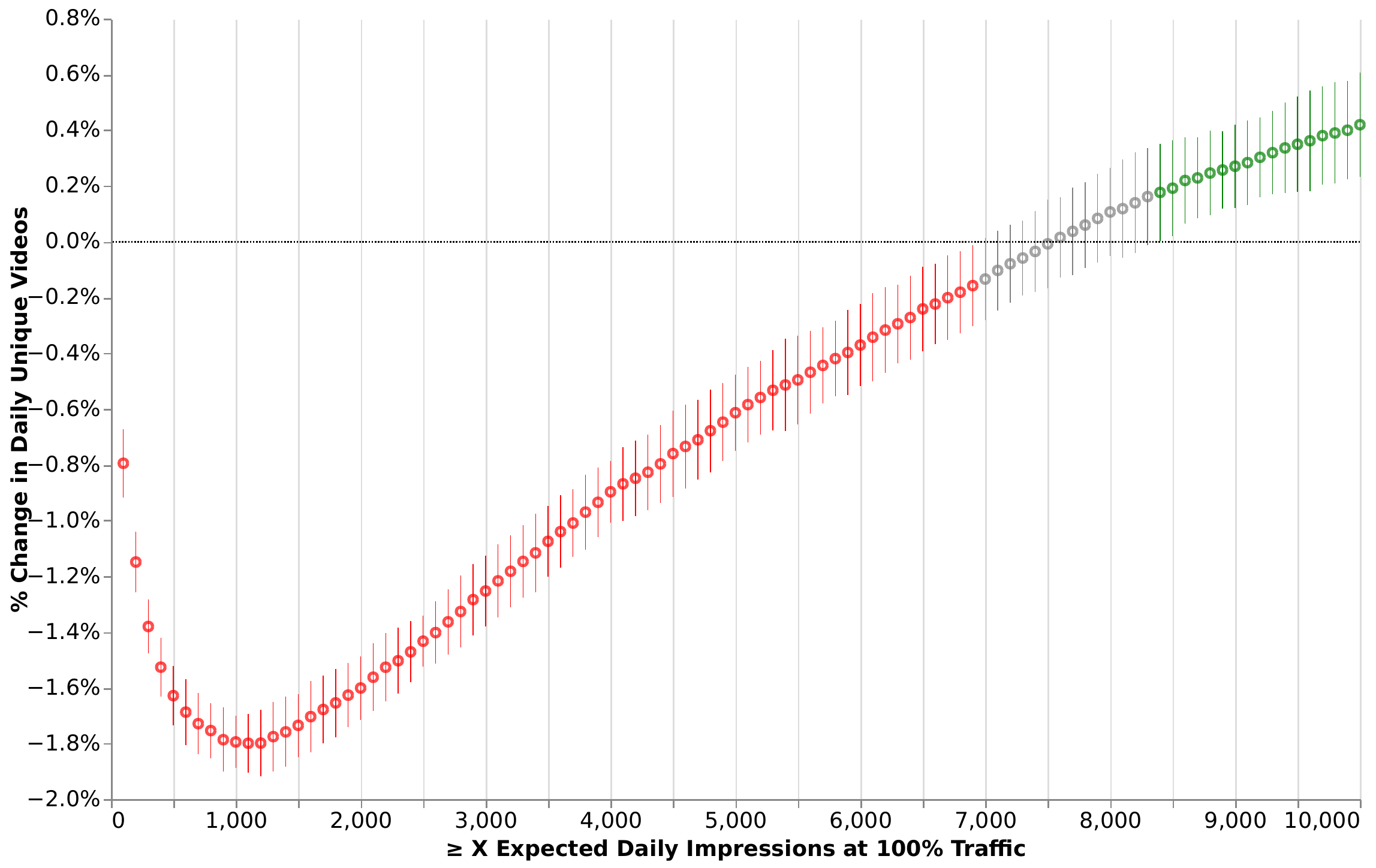}
    }
    \hspace{2cm}
    \subfigure[]
    {
    \includegraphics[width=0.36\textwidth]{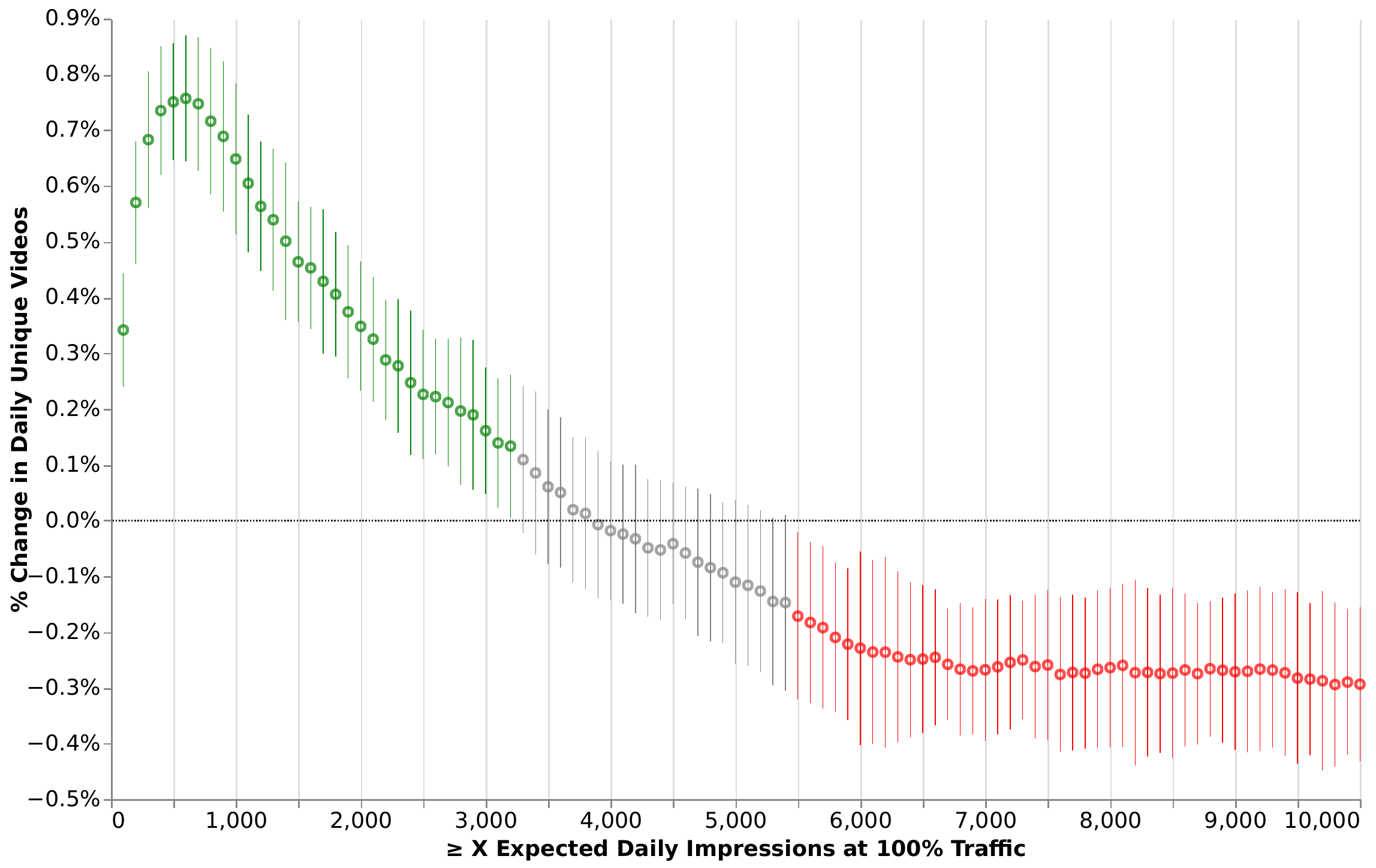}
    }
    }
    \vspace{-0.1in}
    \caption{(left) DUIC@K with 95\% confidence level comparing S-real-time and S-two-tower.  (right) DUIC@K with 95\% confidence level comparing multi-funnel nomination and S-two-tower.}%
    \vspace{-0.05in}
    \label{fig:duiv_multi_funnel}
\end{figure*} 

\begin{table*}
\caption{The improvement of discoverable corpus, i.e., number of contents receiving X post bootstrapping positive interactions  in 7 days compared to S-Two-tower. }
\resizebox{1.0\textwidth}{!}{%
\begin{tabular}{l|llllll}
\hline
X & 100 & 500 & 1K & 2K & 5K & 10K    \\ \hline \hline
S-real-time & \textbf{-1.68\% [-2.33\%, -1.04\%]} & -0.25\% [-1.18\%, 0.68\%] & -0.11\% [-1.22\%, 1.01\%] & -0.49\% [-1.85\%, 0.86\%] & -1.48\% [-3.24\%, 0.29\%] & -1.31\% [-3.42\%, 0.81\%] \\ \hline
Multi-funnel & \textbf{+0.75\% [0.10\%, 1.41\%]} & +0.63\% [-0.36\%, 1.62\%] & \textbf{+1.60\% [0.45\%, 2.75\%]} & \textbf{+1.62\% [0.27\%, 2.96\%]} & +1.63\% [-0.12\%, 3.39\%] & +1.37\% [-0.87\%, 3.61\%] \\ \hline
\end{tabular}%
}
\label{tab:num_contents_7d_post}
\vspace{-0.05in}
\end{table*}
\noindent\textbf{Query Division for Multi-funnel Nomination.} A naive strategy to combine these two nominators is to ask them to nominate candidates separately, and then rely on the graduation filter, prescorer and ranker (in Section~\ref{sec:overview}) to pick the final content for the dedicated slot. But we observe that the middle-funnel contents end up dominating the slot due to undesired popularity bias in the ranker. Activating both nominators for a single user request also incur higher serving cost.  
To mitigate that, we propose query division multiplexing: to randomly select the two-tower DNN to retrieve low-funnel candidates per query with a probability $p\%$ (or real-time nominator to retrieve middle-funnel candidates with probability (100-$p$)\%). We conduct live experiments on different $p$ values for the tradeoff between corpus and user metrics in Section \ref{sec:different_p}.




\section{Experiments}
\label{sec:exp}
In this section, we study the multi-funnel design on its effectiveness in improving the efficiency of dedicated fresh  recommendation stack with a series of live experiments and analyses on a commercial recommendation platform serving billions of users. 

\subsection{Setup}
We test the multi-funnel design in the dedicated fresh recommendation stack as introduced in Section~\ref{sec:overview}. Specifically, we compare the following approaches for dedicated fresh recommendation: 
(i) \textbf{Single-funnel nomination} is used to nominate fresh candidates with a single recommendation model. 
We denote the single-funnel nomination system with the two-tower DNN as S-two-tower, and the one with the real-time sequence model as S-real-time; 
and (ii) \textbf{Multi-funnel nomination} adopts generalization focused two-tower DNN to recommend low-funnel content with under $n_{low}$ positive user interactions\footnote{Note that the S-two-tower in the single-funnel setup uses a different cap that is equal to the graduation threshold, and is much bigger than $n_{low}$}, and the real-time model to recommend middle-funnel content under the graduation threshold. The two nominators are combined through query multiplexing where two-tower DNN is used for $p$\% random user queries while the real-time model for for the rest of $(100-p)$\%. 

We select $p$ to be 80 and  $n_{low}$ to be 200 as discussed in the following Section \ref{sec:different_p}. 
We ran $1$\% online user diverted experiment to measure the user metrics and $5$\%\footnote{We ran corpus user co-diverted A/B testing with a larger traffic since corpus metrics have bigger variance due to the power-law distribution of interactions on contents.} user corpus co-diverted experiment to measure the corpus impact. 



\begin{figure*}
\vspace{-0.1in}
    \graphicspath{{figures/}}
    \centering
    \scalebox{0.93}{
    \subfigure[]
    {
    \includegraphics[width=0.32\textwidth]{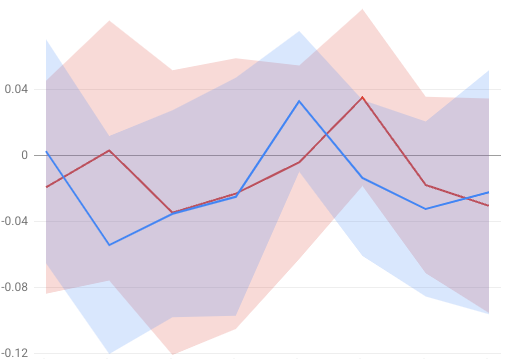}
    }
    \hspace{-0.1cm}
    \subfigure[]
    {
    \includegraphics[width=0.32\textwidth]{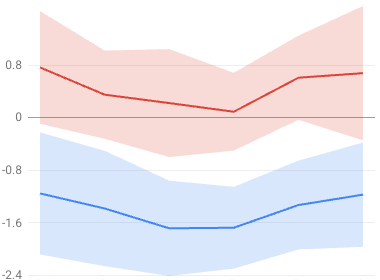}
    }
    \hspace{-0.1cm}
    \subfigure[]
    {
    \hspace{-0.1cm}
    \includegraphics[width=0.32\textwidth]{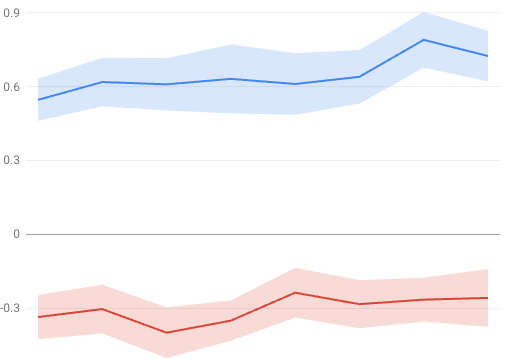}
    }
    }
    \vspace{-0.05in}
    \caption{User metrics on multi-funnel setup: (a) Overall user dwell time on the platform (\% change). (b) User dwell time for small content providers (\% change). (c) Fresh 7 day positive interactions (\% change). Red: multi-funnel; Blue: S-real-time. The baseline is S-Two-tower.}%
    \vspace{-0.05in}
    \label{fig:multi_funnel_user}
\end{figure*} 

\subsection{Performance and Analysis}
\textbf{The impact of multi-funnel nomination.} Comparing the multi-funnel nomination with single-funnel nomination on corpus and user metrics, we make the following observations:
\squishlist
\item \textit{Daily unique impressed contents}. In Figure \ref{fig:duiv_multi_funnel} (left), we find that compared with S-two-tower, S-real-time has significantly lower DUIC at the low end. Specially, it shows 1.79\% degradation at the 1,000 impression threshold, which suggests that the real-time nominator is less efficient than the two-tower DNN model in recommending low-funnel contents with little to no interaction data, leading to significantly lower corpus coverage. 
By combining two-tower DNN to focus on the low-funnel content and the real-time nominator for the middle-funnel content, in Figure \ref{fig:duiv_multi_funnel} (right), we can observe that the DUIC at the low end is significantly improved in the multi-funnel nomination setup, in particular, a 0.65\% improvement of DUIC@1000. This indicates that the multi-funnel recommendation is able to improve the fresh content coverage compared with the single-funnel setup.

\item \textit{Discoverable Corpus}. Table \ref{tab:num_contents_7d_post} summarizes the comparisons among different options. From the results, we find that with the multi-funnel setting, the system is able to improve the metric consistently across the various buckets, and thus enlarges the discoverable corpus. This indicates that multi-funnel nomination is able to identify more quality fresh content that can attract user interactions after exiting the dedicated slot. 
Interestingly, S-real-time does not show statistically significant change for this metric at higher X. We hypothesize that S-real-time alone is not capable of identifying fresh quality content that is also complementary to the main recommendation system. 

\item \textit{User metrics}. 
In Figure \ref{fig:multi_funnel_user}(a), we find that the overall user dwell time on the platform is neutral for both single-funnel (blue) and multi-funnel nomination (red). This suggests that multi-funnel nomination is able to improve corpus metrics without additional cost on short-term user engagement. Furthermore, as observed in Figure \ref{fig:multi_funnel_user}(b), the multi-funnel system shows a $+0.45\%$ increase of the user dwell time for small content providers compared to the single-funnel nominations, demonstrating its strength in uncovering more fresh and tail contents relevant to users. S-real-time, however, shows a significant loss by $-1.41\%$ for that metric, corroborating that it is much less efficient in covering low-funnel corpus. On the other hand, it does achieve a much higher number of fresh 7 day positive interactions (shown in Figure \ref{fig:multi_funnel_user}(c)), suggesting that it infers users' preferences more accurately through utilizing the near real-time interaction feedback. 

\squishend

\begin{table}[h]
\caption{Effect of different maximum positive interaction cap for two-tower DNN. (Compared to positive interaction cap of 2K with the 95\% confidence interval)}
\resizebox{0.46\textwidth}{!}{%
\begin{tabular}{l|lll}
\hline
Cap & 400 & 200 & 100    \\ \hline \hline
DUIC@1K & +0.28\% [0.11\%, 0.46\%] & +0.40\% [0.23\%, 0.58\%] & +0.39\% [0.24\%, 0.54\%] \\ \hline
\end{tabular}%
}
\label{tab:Two-tower DNN_different_viewcap}
\end{table}

\noindent\textbf{The impact of funnel transition cap.} To determine when a fresh content transitions from low funnel to middle funnel, we evaluate the corpus performance of the two-tower DNN for generalization under different interaction caps. Note that when we set the interaction cap to be 100, it means that we limit the corpus indexed by this model to only  fresh contents with maximum interactions of 100. Since the main purpose of low-funnel recommendation is to improve corpus coverage, we mainly focus on the DUIC metrics of different caps, shown in Table~\ref{tab:Two-tower DNN_different_viewcap}. DUIC@1000 reaches its maximum when the cap is set at 200. Setting the cap to 100 achieves similar performance, but further lowering the cap leads to worse metrics. We hypothesize that when the cap is too low, more low quality contents are forced to be nominated by the two-tower model, but later filtered in the ranking stage due to much lower relevance. Indeed, we do observe a 2.9\% drop of the number of unique contents receiving non-zero impressions when the cap is lowed from 400 to 100. Meanwhile, certain amount of initial interactions are needed from the low-funnel nominator to provide learning signals for the real-time model. This suggests a future direction to improve generalization in both the middle funnel nominator as well as main recommendation system such as the ranker, so that the multi-funnel transition can be moved further towards low-funnel. 

\begin{figure}[t]
\includegraphics[width=0.3\textwidth]{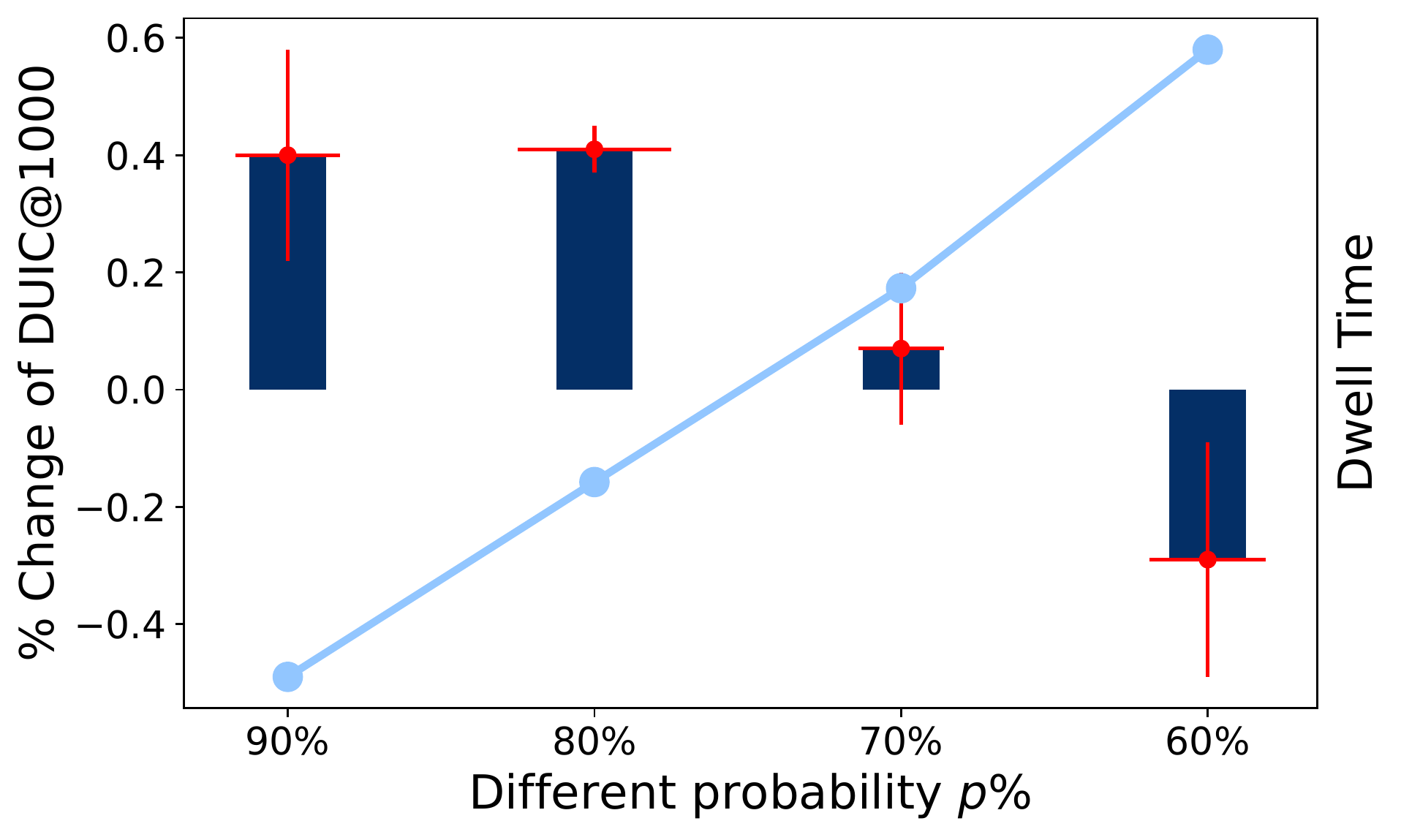}
\caption{Effect of different probability $p\%$. (Left Y axis) Change of DUIC@1000 compared to $p=100$ with the 95\% confidence interval. (Right Y axis) Change in fresh content dwell time (light blue curve). } 
\label{fig:query_differnt_p}
\vspace{-0.05in}
\end{figure}

\smallskip
\noindent\textbf{The impact of different mixing probabilitiy $p\%$.}
\label{sec:different_p}
We also tested different multiplexing probabilities of $p$\% in using two-tower model for nomination (or (100-$p$)\% in using real-time model). As discussed previously, when more random requests adopt the real-time model, users will be matched with more relevant contents due to its near real-time model update. As shown in Figure~\ref{fig:query_differnt_p}, as $p$ decreases, the DUIC gain disappears when $p$\% reduces to 70\%; it, however, increases user dwell time. We select $p$\% to be $80$\% to have a desired trade-off between DUIC@1000 and the user metrics from the dedicated fresh content recommendation. 

\section{Contextual Traffic Assignment}
\label{sec:contextual}
While fresh content recommendation is beneficial for long-term user experience, it comes at the cost of short-term user engagement as less popular or less familiar contents are recommended.  
Users coming to the online platform are often heterogeneous in activity levels, and might have varying levels of interests in consuming such contents. There usually exists a set of \textit{core users} who visit the platform regularly and consistently, while others are \textit{casual} or \textit{emerging} users or who tend to visit the platform occasionally. The heterogeneity in activity levels can lead to distinct content consumption
patterns in different user groups \cite{buttle2004customer,wang2022learning}. And the criteria for grouping the users can be found in \cite{chen2021values}. 

\begin{figure}[t]

\includegraphics[width=0.3\textwidth]{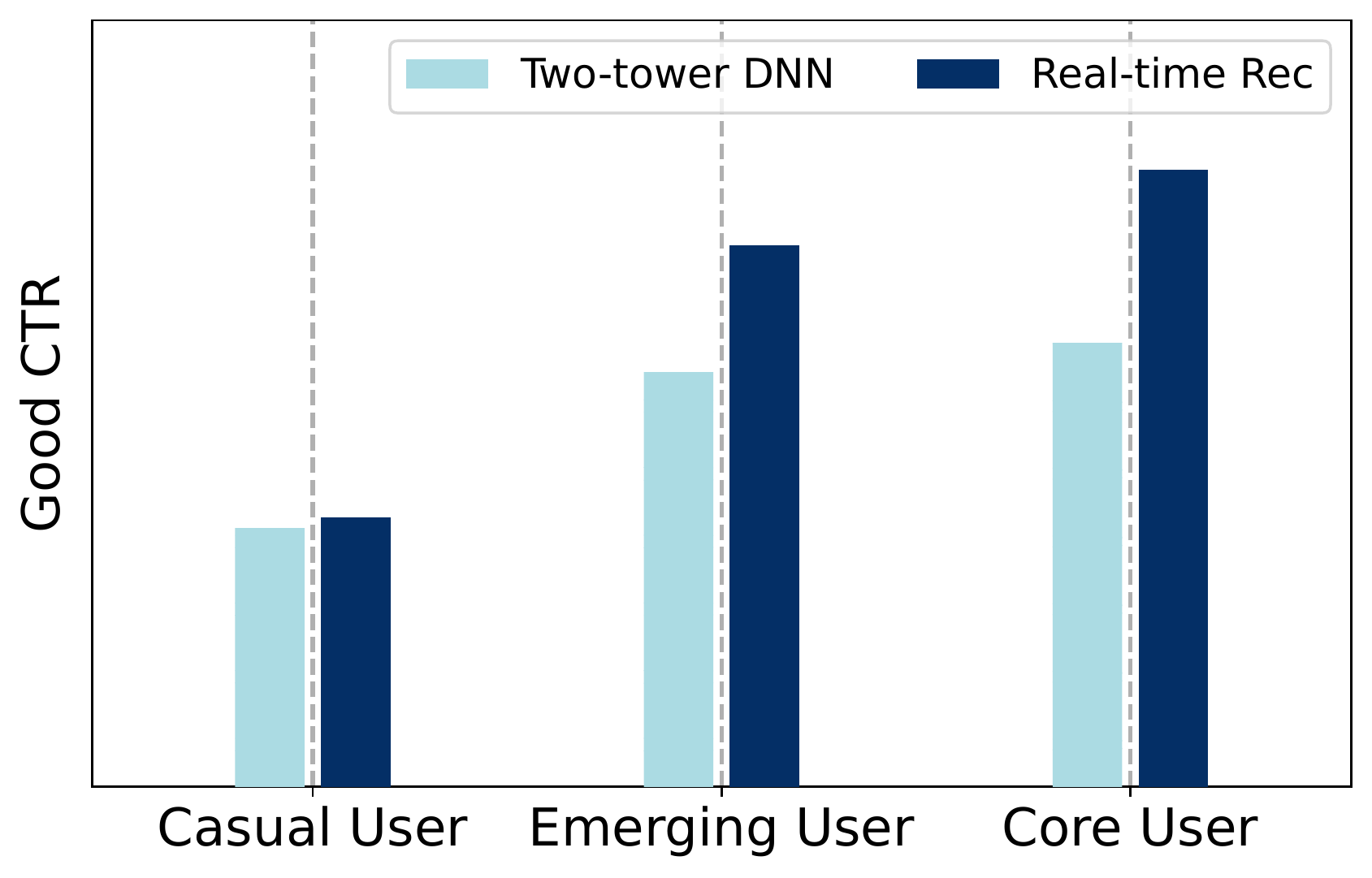}
\vspace{-0.05in}
\caption{The performance of different nominators on users at different activity level. We omit the value of y axis for confidential reason. }
\label{fig:contextual}
\vspace{-0.05in}
\end{figure}

In this initial exploration, we adopt good click-through rate (CTR), conditioning on users spending at least 10 seconds after the click, as a direct user metric to measure the short-term effect of the recommendation. In Figure \ref{fig:contextual}, we find that the good CTR for different user groups are very different on candidates recommended by different models. For example, the low-funnel focused model -- Two-tower DNN achieves similar CTR for casual users compared to the real-time nominator, while it has a significant performance gap on core users.  
This suggests that these models not only have different strength on item corpus, i.e., low-funnel vs middle-funnel, they also have different strength in handling different user groups. 

The analysis motivate a potential approach in further improving the performance of the query multiplexing for multi-funnel nomination. The relevance loss in the generalization model for core users is huge compared to users in lower activity levels. Instead of multiplexing users with different activity levels under the same probability, we can further contextualize the traffic assignment based on users/queries. We chose to randomly select $q\%$ of \textit{core users} and serve them with nominations from the real-time nominator to take advantage of its short-term user engagement gain. Other users are always served with nominations from the two-tower DNN for maximum corpus coverage. In Table \ref{tab:query_differnt_p}, by varying the probability in serving core users with the real-time nominator, we can further improve the recommendation efficiency with context-aware hybrid. For example, when we serve 40\% core users with real-time nominator, we can obtain significant dwell time and good CTR improvement with neutral change in corpus coverage. More sophisticated multiplexing strategies will be investigated.

\section{Related work}
\smallskip
\noindent\textbf{Cold-start Item Recommendation.}
Without enough historic interaction, traditional recommender systems based on collaborative filtering will fail to infer user preference on these newly-available cold-start items~\cite{du2020learn,schnabel2022situating,gope2017survey}. One line of work use the side information, especially  content features of items to compensate the absence of collaborative signals. The core idea  is to approximate
the well-trained collaborative embeddings via content information by modeling their correlation. For example, they try to adopt different objective functions to measure and quantify their correlation, including mean squared root loss \cite{zhu2020recommendation, barkan2019cb2cf,du2020learn}, contrastive loss \cite{wei2021contrastive} and reconstruction loss \cite{zhao2022improving}.
Another line of work solves the cold-start problem by regarding it as a special case of warm-start. For instance, some  \cite{volkovs2017dropoutnet,du2020learn} mimic cold-start by dropping partial training signals for the collaborative embeddings. With the recent advance in meta-learning, the work in \cite{vartak2017meta,pan2019warm,zhu2021learning,lee2019melu} adopt  a learning-to-learn methodology to learn how to adapt the recommender system to newly emerging cold-start items. 
However, most of the existing work in this domain still centers around the offline setups and hard to be deployed on the large-scale real-world systems due to the high latency and resource cost it will introduce. Our work provides practical insights on cold-start item recommendation on industrial content distribution platforms. 
\begin{table}
\centering
\caption{Effect of serving different percentages of core users with real-time nominator, and always serving other users with the two-tower DNN. The baseline is the multi-funnel nomination that randomly serving 20\% users with real-time nominator for all user groups. }
\resizebox{0.48\textwidth}{!}{%
\begin{tabular}{l|llll}
\hline
Core User & DUIC@1K   & num\_contents & dwell\_time & good CTR    \\ \hline
20\% & +0.05\% [-0.13\%, 0.23\%] & +0.364\%&  -1.969\%&  -1.165\%\\
30\% & -0.08\% [-0.25\%, 0.09\%] & -0.385\%&  +3.420\%&  +2.064\%\\
40\% & -0.07\% [-0.25\%, 0.10\%] & -0.163\%&  +8.174\%&  +4.472\%\\ 
50\% & -0.37\% [-0.57\%, -0.17\%] & -3.589\%&  +12.765\%& +7.753\% \\\hline
\end{tabular}%
}
\label{tab:query_differnt_p}
\vspace{-0.1in}
\end{table}

\smallskip
\noindent\textbf{Hybrid Recommendation Systems.} 
In practice, it is impossible to find one model that can obtain the optimal performance in all the scenarios  \cite{aggarwal2016recommender,luo2020metaselector}. 
Serving two or more recommenders simultaneously to take advantages of the strengths of each is widely adopted \cite{burke2002hybrid}. 
Given that collaborative methods are more powerful when abundant data is available and content-based recommendation works better on cold-start items, there are previous discussions on a hybrid setup of combining content-based filtering with collaborative filtering to enable to system to serve both the new and existing users \cite{geetha2018hybrid,jung2004hybrid,ferraro2018automatic}. Early hybridization techniques usually compute a linear combination of individual output scores to blend the results from different recommenders \cite{ekstrand2012recommenders}. With the recent advance in meta-learning, \cite{luo2020metaselector,cunha2016selecting,ding2021learning} propose to meta-learn the hard or soft model selector for the optimal hybridization. Meanwhile, the idea of hybrid recommender systems also have a strong connection to the field of ensemble analysis. In this line of work, instead of combining different recommendation backbones (e.g., content-based method and collaborative filtering), the systems serve multiple models with the same backbone simultaneously \cite{zhang2016ensemble,tang2014ensemble}. We here  take advantage of both the generalization and real-time models in a multi-funnel setup for fresh recommendation.

\section{Conclusion and Future Work}
In this paper, we focus on designing a practical solution to seed the success of high-quality freshly uploaded contents quickly. Specifically, we propose a multi-stage and multi-funnel fresh content recommendation system for a dedicated slot  on a large commercial platform that serves billions of users. We share our lessons in designing a two-tower model with strength in content generalization and a low-latency sequence model consuming the real-time feedback to amplify the worthy content. With live experiments, we demonstrate that combining both models through query multiplexing gains better performance in fresh content recommendation by taking advantages of the strengths of both, and achieves a balance between relevance and coverage. We present a future direction in improving the multi-funnel design: by analyzing the query context, it may further improve the coverage of fresh content recommendation while minimizing the short-term user cost.